\documentclass[11pt,a4paper]{article}

\usepackage[utf8]{inputenc}
\usepackage{amsmath}
\usepackage{amssymb}
\usepackage{graphicx}
\usepackage{hyperref}
\usepackage{natbib}
\usepackage{geometry}
\usepackage{booktabs}
\title{Learning Item Embeddings and Hyperparameters for IRT Calibration via Monte Carlo EM}
\author{James Sharpnack, Kai-Ling Lo\\ Duolingo}
\date{\today}

\begin{document}

\maketitle

\begin{abstract}
Calibration systems for high-stakes computerized adaptive tests (CATs) are essential for growing and maintaining the test's operational item bank.
When a new test item is added to an item bank, we have few response data to provide accurate item parameter estimates, leading to test scores that are poor estimators of the test taker's true score.
Item features and explanatory item response theory (IRT) models mitigate this impact by incorporating item information into the calibration process.
Neural IRT models --- IRT models where the item parameters are the output of a neural net --- provide a powerful framework for accurate calibration, but learning hyperparameters and selecting neural architectures in real time while the CAT is scoring real test takers is impractical and a threat to score validity.
In this work, we propose an initial step prior to launching a CAT that fits a neural net to produce low dimensional embeddings.
The production calibration system can leverage these embeddings to use a simple and interpretable linear explanatory IRT model.
We use a neural parameterization of the 3-parameter logistic (3PL) IRT model in which a feature network maps each item's content features to a low-dimensional representation $h_j = z(x_j) \in \mathbb{R}^d$, from which we model the discrimination and difficulty parameters $(a,b)$ using generalized linear forms.
Due to known identifiability issues when jointly estimating the chance parameter $c$ and test taker ability \citep{lord1980}, we set that to a global constant as opposed to making it also an output of the neural net.
The feature network and the latent test-taker abilities $\theta$ are fit jointly via Monte Carlo Expectation-Maximization (MCEM), removing the need for a separate ability-estimation or pre-calibration stage.
Using an item-split protocol that holds out entire items to simulate feature-only evaluation, we apply this approach to two task types from the Duolingo English Test practice test --- yes/no vocabulary (Y/N Vocab) and vocabulary-in-context (ViC) --- and search over feature sets, network architectures, and representation dimensions $d$.
We find that a shallow two-layer ReLU network with $d=6$ and hand-engineered scalar input features matches or outperforms larger architectures on held-out items for both task types.
This work is a first step toward a compact, content-derived item embedding for use as a feature in the Scalable Parametric Item Calibration Engine (SPICE) \citep{nydick2026scalable}, the fully Bayesian calibration engine at the core of the S2A3 system for adaptive testing \citep{sharpnack2025s2a3}.
\end{abstract}

\section{Introduction}
\label{sec:intro}

Item response theory (IRT) models a test taker's ability and item characteristics, known as item parameters, via parametric models of a test taker's response to a given item \citep{lord1980, wright1979best}.
A key advantage of IRT is interpretability: item parameters can be inspected directly to curate an item bank (e.g., ensuring a wide range of difficulties) and to construct computerized adaptive test (CAT) administration rules.
However, traditional IRT calibration requires many responses per item --- often hundreds --- to meet the standards of a high-stakes test, typically collected during a piloting phase before an item is used for scoring.
Piloting has costs: test takers spend time on items that do not count toward their score; items administered outside the high-stakes test (e.g., on practice tests) are exposed to the public and pose a security risk \citep{laflair2022digital, way1998protecting}; and test takers are less motivated to answer items to the best of their ability on unscored items \citep{cheng2014motivation}.
One way to calibrate new items with few responses is to extend IRT with item features derived from item content, such as NLP features or LLM embeddings \citep{fischer1973lltm, mccarthy2021jump, yancey2024bert}.

A particularly demanding calibration use case is the \emph{feature-only} setting: newly written items are added to the bank with \emph{no} response data, so their item parameters can only be derived from the content of the item itself.
Any feature-only calibration method must ultimately be evaluated by testing on items that are not involved in the training step.  
This is the regime we target throughout this paper, via an item-split protocol (Section~\ref{sec:methods-eval}) that trains the feature-to-parameter mapping on one set of items and evaluates it on a disjoint, held-out set.
The feature-only setting is also the ideal regime for testing learned embeddings, since under this setting, no random effects are needed, making it ideally suited for an initial hyperparameter and neural architecture search stage.

Using a pre-trained neural net to expedite the calibration process has proven fruitful in prior work.
Such blended machine-learning and psychometric methods have been developed at Duolingo under the framework of \emph{computational psychometrics} \citep{vondavier2021computational}, which combines data-driven modeling with the measurement guarantees of psychometrics.
\citet{sharpnack2024autoirt} developed feature-based calibration via automated machine learning (AutoML): a separate model is trained to predict pre-calibrated item parameters from item content, and the IRT model is then refit around these predictions.
Relatedly, \citet{yancey2024bert} fit an explanatory IRT model in which item parameters are linear functions of BERT embeddings and hand-engineered features, accelerating item piloting while retaining IRT interpretability.
However, these works do not learn a neural net embedding, and instead either fit explanatory IRT model directly, or use embeddings from pre-trained foundation models (BERT).
However, none of these works simultaneously learn the test taker abilities as well as item embeddings, which can be used in downstream calibrations.
In this work, we fit the test taker ability, neural net embeddings, and IRT model jointly, in a single end-to-end procedure, and we make the embeddings and hyperparameters computed by this mapping the central object of study.
Concretely, a feature network $z$ maps each item's content features $x_j$ to a $d$-dimensional representation $h_j = z(x_j) \in \mathbb{R}^d$, and the 3PL discrimination, difficulty $(a,b)$ follow a generalized linear form of $h_j$ (Section~\ref{sec:methods-model}).
The 3PL model allows for a chance parameter $c$, which we model as a fixed global constant.
The item parameters are complemented by the test taker ability parameter $\theta$, which provides a score for a test session.
The feature network and generalized-linear item parameter model are fit jointly with the latent test-taker abilities $\theta$ via Monte Carlo Expectation-Maximization (MCEM, Section~\ref{sec:methods-mcem}), which alternates an E-step that samples each test taker's ability from its posterior given the current item-parameter network, and an M-step that updates the network weights by gradient-based maximization of the resulting expected log-likelihood.
This removes the need for a separate ability-estimation or pre-calibration stage: the feature network is trained directly on real response data.

We apply this approach to two task types from the Duolingo English Test practice test --- yes/no vocabulary (Y/N Vocab) and vocabulary-in-context (ViC) --- under the item-split protocol described above, which directly tests whether the learned representation $h_j$ generalizes to unseen items (Section~\ref{sec:methods-eval}).
Section~\ref{sec:results} compares scalar linguistic item features to BERT embeddings (finding that, at the calibration-sample size available here, the high-dimensional embeddings overfit and do not improve over scalar features), searches over feature-network architectures, and asks how small the representation dimension $d$ can be made --- finding that a single shallow architecture with $d=6$ matches or beats higher-dimensional alternatives and deep architectures for both task types --- and examines the guessing parameter $c$ for both task types.
Because the certified DET is comprised of a larger item bank than the practice test, refitting on the larger item bank may render additional BERT features useful.
We use DET practice test in experiment because the certified DET contains sensitive test taker data, however, this methodology will easily generalize to that setting.

\begin{figure}[htbp]
    \centering
    \includegraphics[width=\textwidth]{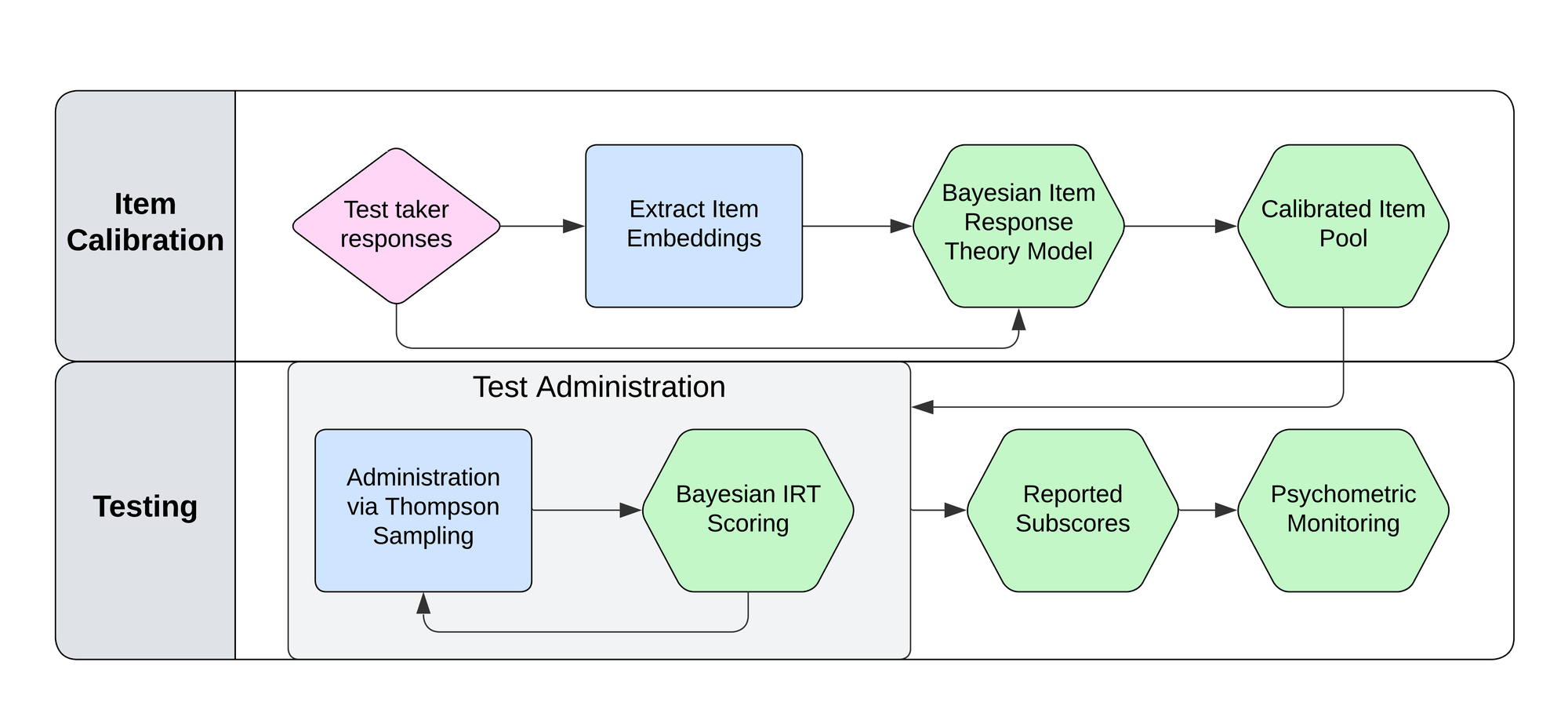}
    \caption{The S2A3 pipeline \citep{sharpnack2025s2a3}: item calibration (top) extracts item
    embeddings from item content, fits a Bayesian IRT model, and
    produces a calibrated item pool; test administration (bottom) uses this pool for
    Thompson-sampling-based item selection, Bayesian IRT scoring, reported subscores, and
    psychometric monitoring. The representation $h_j=z(x_j)$ studied in this paper corresponds
    to the ``Extract Item Embeddings'' step.}
    \label{fig:s2a3-pipeline}
\end{figure}

This work is intended as a first step toward a single, content-derived item embedding for use in the Scalable Parametric Item Calibration Engine (SPICE) \citep{nydick2026scalable}, the fully Bayesian MCMC calibration engine at the core of the S2A3 system for CAT scoring and administration \citep{sharpnack2025s2a3}.
Figure~\ref{fig:s2a3-pipeline} shows this pipeline: item content is used to extract item embeddings, which feed a Bayesian IRT model that produces a calibrated item pool; this pool then drives Thompson-sampling-based test administration, Bayesian IRT scoring, and the resulting subscores and psychometric monitoring.
The representation $h_j = z(x_j)$ studied in this paper comes from the ``Extract Item Embeddings'' step of that pipeline: item content is distilled into a low-dimensional embedding, which is then supplied as a feature to SPICE so that new items --- with few (or no) response data --- contribute to calibration by combining that data with test taker responses on operational item bank, yielding full posterior distributions over $(a,b)$ for use in soft-scoring (S2) and Thompson-sampling-based adaptive administration (A3).
The MCEM procedure of Section~\ref{sec:methods-mcem} is a lightweight, point-estimate analogue of the joint embedding/calibration step that SPICE performs with full posterior uncertainty.

\section{Methodology}
\label{sec:methods}

\subsection{The Neural 3PL Item Response Model}
\label{sec:methods-model}

Let $\theta_i \in \mathbb{R}$ denote the latent ability of test-taker (session) $i$, and let each item $j$ have three IRT parameters: discrimination $a_j > 0$, difficulty $b_j \in \mathbb{R}$, and a guessing/pseudo-chance parameter $c_j \in [0,1)$.
The probability that session $i$ answers item $j$ correctly under the 3-parameter logistic (3PL) model is
\begin{equation}
\label{eq:3pl}
P(Y_{ij}=1 \mid \theta_i) = c_j + (1 - c_j)\,\sigma\big(a_j(\theta_i - b_j)\big),
\end{equation}
where $\sigma(x) = 1/(1+\exp(-x))$ is the logistic function.
Setting $c_j \equiv 0$ recovers the 2PL model.
Recall that in our model we treat $c_j$ as a hyperparameter, with either $c_j = c$ (a global constant) or a simple heuristic, such as different values for course groupings of items.

Rather than treating $(a_j, b_j)$ as free per-item parameters (the classical IRT approach, also implemented here as a baseline with one learned $(a,b)$ pair per item), we predict them from a feature vector $x_j \in \mathbb{R}^F$ describing item $j$ (e.g.\ word frequency statistics, length, CEFR level, the \texttt{is\_real} indicator, etc.) via a feed-forward network.
A feature multi-layer perception ($\mathrm{MLP}$) produces a $d$-dimensional representation
\begin{equation}
    \label{eq:embeddingmapping}
h_j = \mathrm{MLP}(x_j) \in \mathbb{R}^d,
\end{equation}
which is mapped to the $(a,b)$ parameters via, $(\lambda_{a_j}, \lambda_{b_j}) = W h_j + u$, with
\begin{equation}
    \label{eq:embeddingirtmapping}
a_j = \exp(\lambda_{a_j}) \qquad b_j = \lambda_{b_j}.
\end{equation}
The exponential (log-linear) link guarantees $a_j > 0$ and is our chosen parameterization for the IRT discrimination.
Out of concern that the exponential form would cause extreme discrimination values, we also attempted other parametrizations with linear growth.
A softplus link $a_j = \mathrm{softplus}(\lambda_{a_j})$ was also tried; Section~\ref{sec:disc-link} compares the two and finds them equivalent under item-split evaluation.
A bounded alternative, $a_j = a_{\text{cap}}\,\sigma(\lambda_{a_j})$ with a fixed cap $a_{\text{cap}}$, is also supported but not used by default.
Section~\ref{sec:results} searches over the depth, width, and activation of this MLP under the constraint $d=6$.

For the guessing parameter $c_j$, three variants were compared for the Y/N Vocab item type: (i) a single global constant $\texttt{fixed\_c}$, shared by all items and fixed by hand rather than learned; (ii) an item-type-dependent pair $(c_{\text{real}}, c_{\text{fake}})$ selected via an item-level ``is real'' indicator, $c_j = c_{\text{real}}$ if item $j$ is ``real'' and $c_j = c_{\text{fake}}$ otherwise (used for Y/N Vocab, where ``fake'' items are plausible non-words with a structural guessing floor).
In all experiments reported here we use a \emph{fixed} $c$ --- variant (i) for ViC ($c=0$) and the item-type pair (ii) for Y/N Vocab --- and select its value by grid search over held-out log-loss (Section~\ref{sec:stv-c-grid}).

\subsection{Item Types}
\label{sec:methods-item-types}

We study the first two item types administered in the Duolingo English Test (DET) practice test: yes/no vocabulary (Y/N Vocab) and vocabulary-in-context (ViC) \citep{cardwell2022duolingo}.
In Y/N Vocab, the test-taker is shown a single string and must decide whether it is a real English word.
The items are drawn from a pool of real words and plausible-looking fake words (the fake words are generated by a recurrent neural network), so a correct response can be obtained by guessing the yes/no decision; this gives the task a structural guessing floor that motivates a nonzero pseudo-chance parameter $c$ for fake items.
In ViC, the test-taker completes a partially-revealed word embedded in a sentence that supplies context: a visible prefix is shown and the remaining characters are blanked (e.g.\ ``inter\rule{1.2em}{0.4pt}''), and the test-taker must type the missing characters.
Because reproducing an exact character sequence is far harder to guess than a binary decision, ViC has no comparable guessing floor and we set $c=0$.

\subsection{Item Features}
\label{sec:methods-features}

The feature vector $x_j$ is computed from the content of item $j$ alone, so that it is available even for a unseen item with no responses.
For features derived from a word's usage in a corpus, we use the Corpus of Contemporary American English (COCA) \citep{davies2008word}, which is partitioned into eight genre sub-corpora (blog, web, TV/movie, spoken, fiction, magazine, news, and academic).
The two task types we study --- yes/no vocabulary (Y/N Vocab) and vocabulary-in-context (ViC) --- have different content, and so use different feature sets, described below.

For Y/N Vocab, an item consists of a single (real or fake) word, and we use $46$ features.
Some are evident from the word in isolation: the binary \texttt{is\_real} indicator (true for words found in COCA, false for plausible generated non-words), the word length in characters, and binary indicators of whether the word appears in each CEFR level's wordlist (A1 through C2) \citep{councilofeurope2001}.
Several features summarize the word's corpus usage: its log frequency and log frequency-rank in COCA, and the proportion of its occurrences that are capitalized.
We also include the three smallest Levenshtein (edit) distances from the word to any real word, which are all zero for real words but locate fake words relative to the real vocabulary.
Finally, we include a family of character n-gram features computed from COCA: for each $n$ from $1$ to $4$, the frequency and frequency-rank of the word's length-$n$ prefix and suffix, the proportion of the word's length-$n$ character n-grams whose frequency exceeds a fixed threshold, and the minimum, mean, and maximum log frequency-rank over the word's length-$n$ n-grams.
These features capture how typical the word's sub-word spelling patterns are in English, which is especially informative for distinguishing fake words.

For ViC, an item is a passage with a damaged target word, and we use $15$ scalar features.
Surface features describe the target itself: its length in characters and the percentage of its characters that are vowels.
We also include a relative damaged-probability feature: the COCA frequency of the true target relative to the other real words consistent with the visible (undamaged) letters, which measures how recoverable the target is from the surviving characters.
Context-frequency features describe how common the target word is: its Laplace-smoothed log frequency in COCA overall, its log frequency per million in each of the eight COCA sub-corpora, and its log \emph{document} frequency (the number of COCA documents in which it appears).
We further include the average log word frequency over the entire passage and the normalized position of the damaged word within the passage (its token index divided by the passage length).
Unlike AutoIRT \citep{sharpnack2024autoirt} and BERT-IRT \citep{yancey2024bert}, which incorporate contextual BERT embeddings of the passage as additional features, our default ViC configuration uses only these scalar features; Section~\ref{sec:vic-embeddings} shows that, with the limited number of calibration items available here, adding BERT embeddings degrades held-out accuracy --- the high-dimensional embeddings overfit the item sample rather than adding usable signal --- though a larger item bank could change this.

All of these features are extracted programmatically from the item content by a deterministic featurizer, with no human annotation in the loop.
For Y/N Vocab, the word- and character-level statistics are obtained by table lookups against the COCA frequency and rank tables and the CEFR wordlists; for ViC, the target and passage are scored by analogous word-, sub-corpus-, and document-frequency lookups.
Because the featurizer consumes only the item text --- and no response data --- it applies unchanged to a newly written item, which is precisely what makes content-derived feature-only calibration possible.
This connects naturally to automated item generation (AIG), which has been used at Duolingo to generate items such as interactive reading passages and their comprehension questions \citep{attali2022interactive}: the Y/N Vocab fake words are themselves produced by a recurrent neural network (Section~\ref{sec:methods-item-types}), and because calibration here depends only on content-derived features, the same pipeline can featurize and calibrate automatically generated items at scale --- before any test taker has responded to them.

\subsection{Monte Carlo EM}
\label{sec:methods-mcem}

The abilities $\theta_i$ are latent variables.
We fit the network parameters $\phi$ (the weights of the feature MLP and $(a,b)$ head) by Monte Carlo EM, alternating an E-step that samples $\theta_i$ given the current $\phi$ and an M-step that updates $\phi$ given the sampled $\theta_i$.

\paragraph{E-step.} With $\phi$ fixed, the posterior over $\theta_i$ given session $i$'s responses $y_i = (y_{ij})_{j \in \text{session } i}$ is
\begin{equation}
p(\theta_i \mid y_i, \phi) \;\propto\; p(\theta_i) \prod_{j \in \text{session } i} P(Y_{ij}=y_{ij} \mid \theta_i, \phi),
\end{equation}
with a $\mathcal{N}(0,1)$ prior $p(\theta)$.
We approximate this posterior on a discrete grid $\{\theta_g\}_{g=1}^G$ (default $G=60$ points on $[-3,3]$): for each grid point we sum the per-observation log-likelihoods (Eq.~\ref{eq:3pl}) across the session, add the log prior, and normalize with a softmax over $g$ to obtain $p(\theta_g \mid y_i)$.
$K$ independent draws $\theta_i^{(1)},\dots,\theta_i^{(K)}$ are then obtained per session by categorical sampling over the grid, each perturbed by uniform jitter in $[-\Delta/2, \Delta/2]$ (where $\Delta$ is the grid spacing) to avoid discretization artifacts.

\paragraph{M-step.} Given the $K$ draws per session, $\phi$ is updated by the Adam optimizer \citep{kingma2014adam} to maximize the Monte Carlo estimate of the expected complete-data log-likelihood,
\begin{equation}
Q(\phi) = \frac{1}{K} \sum_{k=1}^{K} \sum_{i} \sum_{j \in \text{session } i} \log P\big(Y_{ij}=y_{ij} \mid \theta_i^{(k)}, \phi\big),
\end{equation}
running \texttt{n\_epochs} full-batch gradient steps.
$K=1$ corresponds to ordinary stochastic EM; for $K>1$ each session contributes $K$ independent copies to the loss.

\paragraph{Initialization and schedule.} Rather than initializing $\theta_i^{(0)}$ from the prior, we initialize it from an external ability proxy ($\theta_{\text{proxy}}$, e.g.\ the logit of each session's empirical grade rate), which prior tuning confirmed speeds convergence relative to a random or prior-drawn initialization.
The E/M steps then alternate for $T$ iterations.
Unless otherwise noted, experiments use $T=30$, $K=1$, $50$ epochs, and the Adam optimizer with learning rate of $10^{-3}$, all established in earlier hyperparameter tuning on the Y/N Vocab data.

\subsection{Evaluation Protocol}
\label{sec:methods-eval}

The primary evaluation metric is held-out binary cross-entropy (log-loss) on responses to items not used to fit the feature mapping, $x\to(a,b)$.
Sections~\ref{sec:results} onward use an \emph{item-split} protocol (Section~\ref{sec:item-split-data}): entire items, rather than sessions or individual responses, are held out, directly testing whether the learned mapping generalizes to unseen items.
Models are evaluated once on the held-out set, and bootstrap $95\%$ confidence intervals on the mean log-loss are obtained from $200$ resamples of the held-out predictions.

\section{Results}
\label{sec:results}

We study two item types from the Duolingo English Test practice test --- the Y/N Vocab dataset and the Vocabulary-in-Context (ViC) dataset --- under a common item-split evaluation protocol.
The response data come from the free, low-stakes practice test rather than the certified, high-stakes Duolingo English Test.
For both datasets we ask whether a low-dimensional representation of item features, fixed at $6$ dimensions, can be mapped through a small MLP to the discrimination and difficulty parameters $(a,b)$ while achieving both low held-out log-loss and a meaningfully wide, structured discrimination distribution.
``Item-split'' means that entire items --- not individual sessions or responses --- are held out, directly testing whether the learned feature mapping, $x_j \to h_j$ \eqref{eq:embeddingmapping}, generalizes to items unseen during training.

\subsection{Item-Split Datasets}
\label{sec:item-split-data}

For each dataset we (1) shuffle the unique item identifiers; (2) hold out $20\%$ of items as a test set, with the remaining $80\%$ as train items; (3) subsample each item's responses to at most $1000$, sampled without replacement (seed $42$ for train items, seed $43$ for test items), so that no single high-volume item dominates the loss; and (4) train on the existing $\theta_{\text{proxy}}$ ability estimates for train-item responses, evaluating held-out log-loss on test-item responses.
Table~\ref{tab:item-split-sizes} summarizes the resulting datasets.

\begin{table}[h]
\centering
\scriptsize
\caption{Item-split dataset sizes for ViC and Y/N Vocab.}
\label{tab:item-split-sizes}
\begin{tabular}{lrrrrrrr}
\toprule
Dataset & Total items & Train items & Test items & Train rows & Test rows & Resp./item (train) & Resp./item (test) \\
\midrule
ViC & 585 & 468 & 117 & 468{,}000 & 117{,}000 & 1000.0 & 1000.0 \\
Y/N Vocab & 3326 & 2661 & 665 & 1{,}210{,}026 & 306{,}309 & 454.7 & 460.6 \\
\bottomrule
\end{tabular}
\end{table}

Y/N Vocab responses per item are heavily right-skewed (mean $\approx 600$, median $\approx 381$, max $\approx 9805$), so the cap mainly trims a long tail and the post-subsampling average ($\approx 455$--$461$ responses/item) remains well below $1000$.
Both datasets now share a directly comparable item-split harness; all subsequent results in this section train on item-split train data and report held-out log-loss, with bootstrap $95\%$ confidence intervals (200 resamples), on the held-out items.

\subsection{Discrimination Link Function}
\label{sec:disc-link}

Before the representation study, we fix the link that maps the network's raw discrimination output $\lambda_{a_j}$ to the positive discrimination $a_j$.
We compare the exponential link $a_j = \exp(\lambda_{a_j})$ against the softplus link $a_j = \mathrm{softplus}(\lambda_{a_j})$ on the Y/N Vocab data, using the item-split protocol (Section~\ref{sec:item-split-data}) and a fixed architecture ($46$ features $\to$ \texttt{Linear(46,6)} $\to$ identity $\to$ \texttt{Linear(6,2)}, with $c=0.25$ for every item), trained for $20$ epochs and evaluated over seeds $\{42, 123\}$ to bracket the seed-noise floor.
Table~\ref{tab:disc-link} reports held-out log-loss (with bootstrap $95\%$ CIs) and the maximum fitted discrimination $a_{\max}$.

\begin{table}[h]
\centering
\caption{Discrimination link comparison on Y/N Vocab item-split data ($c=0.25$).}
\label{tab:disc-link}
\begin{tabular}{llccc}
\toprule
Link & Seed & val. loss & 95\% CI & $a_{\max}$ \\
\midrule
exp & 42 & 0.51388 & [0.51191, 0.51600] & 1.02 \\
exp & 123 & 0.51185 & [0.50981, 0.51388] & 1.15 \\
softplus & 42 & 0.51367 & [0.51169, 0.51579] & 0.88 \\
softplus & 123 & 0.51197 & [0.50993, 0.51402] & 0.94 \\
\bottomrule
\end{tabular}
\end{table}

Under the item-split protocol the two links are statistically indistinguishable: at matched seed the held-out log-loss differs by at most $\approx 0.0002$ --- an order of magnitude below the seed-to-seed variation of $\approx 0.0017$ --- and all bootstrap CIs overlap heavily.
Both links are also numerically stable here, with $a_{\max} \leq 1.15$ for every configuration; the representation search of Section~\ref{sec:repr6}, run under the exponential link, likewise shows no discrimination blow-ups across any architecture.
We adopt the exponential (log-linear) link as the default throughout the remaining experiments: it is the standard IRT discrimination parameterization, it is marginally the better of the two across architectures on Y/N Vocab (Section~\ref{sec:repr6}), and under item-split evaluation it never exhibits the unbounded growth that the unconstrained exponential can in principle produce.

\subsection{ViC: Scalar Features versus BERT Embeddings}
\label{sec:vic-embeddings}

For ViC we first ask whether contextual BERT embeddings of the blanked target position in the passage (the \emph{mask} embedding) and of the candidate key/answer word (the \emph{key} embedding) add information beyond a set of $15$ hand-engineered scalar item features (word frequency measures, key length, vowel percentage, part-of-speech ratios, etc.).
All variants use a simple generalized linear model (single linear hidden layer in the neural net), with weight decay $10^{-2}$, and $c=0$ (a fill-in-the-blank task has no guessing floor).
Regularization approaches such as $L_1$ penalization and dropout layers were attempted with nearly identical results.
Table~\ref{tab:vic-embeddings} reports held-out log-loss for each input variant.

\begin{table}[h]
\centering
\footnotesize
\caption{ViC item-split log-loss by input feature variant.}
\label{tab:vic-embeddings}
\begin{tabular}{lccl}
\toprule
Variant & val. loss & 95\% CI & Description \\
\midrule
\textbf{scalars} & \textbf{0.4309} & [0.4280, 0.4336] & 15 scalar item features only \\
scalars + mask + key & 0.4779 & [0.4738, 0.4816] & scalars plus mask and key BERT embeddings \\
scalars + mask + key + cosine & 0.4780 & [0.4740, 0.4817] & previous row plus $\cos(\text{mask}, \text{key})$ \\
scalars + key & 0.5231 & [0.5193, 0.5274] & scalars plus key BERT embedding \\
scalars + mask & 0.5389 & [0.5344, 0.5439] & scalars plus mask BERT embedding \\
embeddings only & 0.5393 & [0.5345, 0.5438] & mask and key BERT embeddings, no scalars \\
\bottomrule
\end{tabular}
\end{table}

The $15$ scalar features alone (the \emph{scalars} variant) outperform every BERT-embedding variant by a wide margin ($\geq 0.04$ in val.\ loss, far outside any CI overlap).
Adding BERT embeddings on top of the scalar features actively hurts held-out performance, and using the embeddings alone (no scalars) is worst of all.
We read this not as evidence that BERT embeddings are intrinsically uninformative for this task, but as a sample-size effect: the calibration set here --- a few hundred ViC items --- is too small to fit the high-dimensional embedding-to-parameter mapping without overfitting, so the added capacity is spent memorizing training items rather than generalizing to held-out ones.
We therefore use scalar features only, with no embedding tower, for ViC in the remaining experiments, while noting that a substantially larger item bank could make BERT features worthwhile.

\subsection{ViC: Scalar-Feature MLP Architecture Search}
\label{sec:vic-mlp}

Using the $15$ scalar features as input, we compare $10$ MLP architectures (varying depth, width, activation, and dropout) feeding an IRT head, again with weight decay $10^{-2}$, and $c=0$, averaged over seeds $\{42, 123\}$.
We name each architecture by its number of hidden layers, activation, and hidden-layer widths: for example, \texttt{1L\_relu\_16} is a single ReLU hidden layer of width $16$, \texttt{2L\_relu\_16\_8} stacks two ReLU layers of widths $16$ then $8$, and \texttt{3L\_relu\_16\_8\_4} stacks three (widths $16$, $8$, $4$); a \texttt{\_dropout} suffix adds dropout with rate $0.2$ (in the fixed-$6$-dimensional search of Section~\ref{sec:repr6} the final hidden width is always $6$, e.g.\ \texttt{2L\_relu\_8\_6} for widths $8$ then $6$).
Table~\ref{tab:vic-mlp} reports held-out log-loss together with summary statistics of the fitted discrimination parameter $a_j$ taken over items: the mean $a_{\text{mean}}$, standard deviation $a_{\text{std}}$, $95$th percentile $a_{p95}$, and maximum $a_{\max}$.

\begin{table}[h]
\centering
\caption{ViC scalar-feature MLP architecture search (mean over seeds 42, 123).}
\label{tab:vic-mlp}
\begin{tabular}{lccccc}
\toprule
Architecture & val. loss & $a_{\text{mean}}$ & $a_{\text{std}}$ & $a_{p95}$ & $a_{\max}$ \\
\midrule
1L\_relu\_16 & 0.4306 & 1.234 & 0.141 & 1.457 & 1.563 \\
1L\_relu\_8 & 0.4308 & 1.234 & 0.144 & 1.457 & 1.570 \\
1L\_identity\_8 & 0.4309 & 1.173 & 0.098 & 1.353 & 1.461 \\
1L\_relu\_4 & 0.4310 & 1.220 & 0.146 & 1.473 & 1.627 \\
2L\_relu\_16\_8 & 0.4310 & 1.249 & 0.155 & 1.580 & 1.670 \\
3L\_relu\_16\_8\_4 & 0.4310 & 1.253 & 0.145 & 1.539 & 1.613 \\
1L\_tanh\_8 & 0.4312 & 1.189 & 0.102 & 1.357 & 1.440 \\
2L\_relu\_8\_4 & 0.4313 & 1.244 & 0.153 & 1.570 & 1.652 \\
2L\_relu\_16\_8\_dropout & 0.4314 & 1.202 & 0.124 & 1.408 & 1.454 \\
1L\_sigmoid\_8 & 0.4344 & 1.221 & 0.107 & 1.412 & 1.471 \\
\bottomrule
\end{tabular}
\end{table}

ReLU architectures (val.\ loss $0.4306$--$0.4313$) and the identity activation ($0.4309$) are statistically tied for lowest test set loss.
The more salient difference is in the spread of the discrimination parameter: ReLU's $a_{\text{std}}$ ($0.14$--$0.16$) is markedly larger than identity's ($0.10$), with $a_{\max}$ reaching $\approx 1.67$ versus $\approx 1.46$.
To check whether this wider spread reflects a real, stable signal or training noise, we compute per-item $(a,b)$ values (independent of $\theta$) for four representative architectures and compare seeds $42$ and $123$, and correlate the average $a$ with item-level $\log$-frequency (Table~\ref{tab:vic-stability}).

\begin{table}[h]
\centering
\footnotesize
\caption{Seed stability of per-item $(a,b)$ and correlation with word frequency.}
\label{tab:vic-stability}
\begin{tabular}{lccc}
\toprule
Architecture & $a$ corr.\ (Pearson / Spearman) & $b$ corr.\ (Pearson / Spearman) & corr.$(\bar{a}, \log\text{-freq})$ \\
\midrule
1L\_identity\_8 & 0.996 / 0.995 & 1.000 / 1.000 & $-0.620$ \\
1L\_relu\_8 & 0.984 / 0.986 & 0.999 / 0.997 & $-0.676$ \\
1L\_relu\_16 & 0.999 / 0.998 & 0.999 / 0.999 & $-0.666$ \\
3L\_relu\_16\_8\_4 & 0.999 / 0.998 & 0.999 / 0.998 & $-0.910$ \\
\bottomrule
\end{tabular}
\end{table}

The seed-to-seed correlations of per-item $a$ and $b$ are $\geq 0.984$ (mostly $\geq 0.996$) for every architecture, and the wider ReLU spread is strongly tied to word frequency (correlation with $\log$-frequency as strong as $-0.91$ for the three-layer ReLU network, versus $-0.62$ for identity).
This indicates that the wider discrimination spread produced by ReLU architectures is a real, reproducible signal --- rarer words receive higher estimated discrimination --- rather than an artifact of training noise.
Sigmoid is clearly the worst activation ($0.4344$).
We conclude that ReLU-family architectures are preferable for ViC.

\subsection{Y/N Vocab: Guessing-Parameter Grid Search on Item-Split Data}
\label{sec:stv-c-grid}

For Y/N Vocab, items carry an \texttt{is\_real} flag distinguishing real words from plausible-looking fake words, and we allow separate guessing parameters $c_{\text{real}}$ and $c_{\text{fake}}$, with $c = c_{\text{real}}$ if \texttt{is\_real} and $c = c_{\text{fake}}$ otherwise.
With the architecture fixed at $46$ input features ($45$ scalar features plus \texttt{is\_real}, $z$-scored) $\to$ \texttt{Linear(46,6)} $\to$ Sigmoid $\to$ \texttt{Linear(6,2)} (the head producing $a,b$), weight decay $10^{-2}$, seed $42$, we grid-search $c_{\text{real}}, c_{\text{fake}} \in \{0.0, 0.1, 0.2, 0.3, 0.4, 0.5\}$ ($36$ combinations) on the item-split data.
Table~\ref{tab:stv-cgrid-top} shows the eight best combinations by held-out log-loss.

\begin{table}[h]
\centering
\caption{Top eight $(c_{\text{real}}, c_{\text{fake}})$ combinations on Y/N Vocab item-split data.}
\label{tab:stv-cgrid-top}
\begin{tabular}{ccccc}
\toprule
$c_{\text{real}}$ & $c_{\text{fake}}$ & val. loss & 95\% CI & $a_{\text{mean}}$ ($a_{\max}$) \\
\midrule
\textbf{0.1} & \textbf{0.0} & \textbf{0.51338} & [0.51134, 0.51535] & 0.242 (0.429) \\
0.2 & 0.0 & 0.51339 & [0.51135, 0.51536] & 0.234 (0.393) \\
0.0 & 0.0 & 0.51357 & [0.51152, 0.51556] & 0.249 (0.460) \\
0.3 & 0.0 & 0.51370 & [0.51167, 0.51571] & 0.224 (0.351) \\
0.1 & 0.1 & 0.51417 & [0.51213, 0.51615] & 0.232 (0.440) \\
0.2 & 0.1 & 0.51418 & [0.51214, 0.51616] & 0.223 (0.403) \\
0.0 & 0.1 & 0.51437 & [0.51232, 0.51636] & 0.239 (0.471) \\
0.3 & 0.1 & 0.51449 & [0.51245, 0.51649] & 0.213 (0.360) \\
\bottomrule
\end{tabular}
\end{table}

Two one-dimensional slices through the grid (Table~\ref{tab:stv-cgrid-slices}) show that, within every $c_{\text{real}}$ row, held-out log-loss rises monotonically with $c_{\text{fake}}$, and $c_{\text{fake}}=0$ is best for every value of $c_{\text{real}}$.

\begin{table}[h]
\centering
\caption{One-dimensional slices through the Y/N Vocab $(c_{\text{real}}, c_{\text{fake}})$ grid.}
\label{tab:stv-cgrid-slices}
\begin{minipage}{0.48\textwidth}
\centering
\begin{tabular}{cccc}
\toprule
$c_{\text{real}}$ ($c_{\text{fake}}=0$) & val. loss & $a_{\text{mean}}$ & $a_{\max}$ \\
\midrule
0.0 & 0.51357 & 0.249 & 0.460 \\
0.1 & 0.51338 & 0.242 & 0.429 \\
0.2 & 0.51339 & 0.234 & 0.393 \\
0.3 & 0.51370 & 0.224 & 0.351 \\
0.4 & 0.51450 & 0.212 & 0.342 \\
0.5 & 0.51613 & 0.197 & 0.350 \\
\bottomrule
\end{tabular}
\end{minipage}
\hfill
\begin{minipage}{0.48\textwidth}
\centering
\begin{tabular}{cccc}
\toprule
$c_{\text{fake}}$ ($c_{\text{real}}=0$) & val. loss & $a_{\text{mean}}$ & $a_{\max}$ \\
\midrule
0.0 & 0.51357 & 0.249 & 0.460 \\
0.1 & 0.51437 & 0.239 & 0.471 \\
0.2 & 0.51553 & 0.227 & 0.483 \\
0.3 & 0.51713 & 0.212 & 0.496 \\
0.4 & 0.51935 & 0.195 & 0.511 \\
0.5 & 0.52296 & 0.177 & 0.531 \\
\bottomrule
\end{tabular}
\end{minipage}
\end{table}

The combinations $(c_{\text{real}}, c_{\text{fake}}) \in \{0.0,0.1,0.2,0.3\} \times \{0.0\}$ are essentially tied (within $\approx 0.0003$ val.\ loss of one another), with the overall best at $c_{\text{real}}=0.1, c_{\text{fake}}=0.0$ (val.\ loss $0.5134$).
This contrasts with our earlier reliability-focused tuning, which favored $c_{\text{fake}}=0.5$ under a within-session split-half reliability objective; the two studies optimize different quantities (split-half reliability versus item-split held-out log-loss) and are not directly comparable.
We fix $c_{\text{real}}=0.1, c_{\text{fake}}=0.0$ for the representation search below.
However, it should be noted that the log-likehood is not sensitive to choice of $c$, particularly when it does not exceed $0.4$.

\subsection{Fixed Six-Dimensional Representation Search}
\label{sec:repr6}

Finally, for both datasets we search over MLP architectures whose final hidden layer --- the representation passed to the $(a,b)$ head --- is fixed at $6$ dimensions, reflecting an ``at most $8$-dimensional, hence $6$-dimensional'' budget for the learned representation.
Both datasets share a model class in which $c = c_{\text{real}}$ for ``real'' items and $c = c_{\text{fake}}$ otherwise; for ViC every item is treated as ``real'' with $c_{\text{real}}=c_{\text{fake}}=0$, and for Y/N Vocab we use $c_{\text{real}}=0.1, c_{\text{fake}}=0.0$ from Section~\ref{sec:stv-c-grid}.
Table~\ref{tab:repr6-archs} lists the nine architectures considered, all of which terminate in a $6$-dimensional layer feeding \texttt{Linear(6,2)}.

\begin{table}[h]
\centering
\caption{Architectures for the fixed 6-dimensional representation search.}
\label{tab:repr6-archs}
\begin{tabular}{lccc}
\toprule
Name & Hidden dims & Activation & Dropout \\
\midrule
1L\_identity\_6 & [6] & identity & 0.0 \\
1L\_relu\_6 & [6] & relu & 0.0 \\
1L\_sigmoid\_6 & [6] & sigmoid & 0.0 \\
1L\_tanh\_6 & [6] & tanh & 0.0 \\
2L\_relu\_8\_6 & [8, 6] & relu & 0.0 \\
2L\_relu\_12\_6 & [12, 6] & relu & 0.0 \\
2L\_sigmoid\_8\_6 & [8, 6] & sigmoid & 0.0 \\
3L\_relu\_16\_8\_6 & [16, 8, 6] & relu & 0.0 \\
2L\_relu\_8\_6\_dropout & [8, 6] & relu & 0.2 \\
\bottomrule
\end{tabular}
\end{table}

All configurations use weight decay $10^{-2}$, and seeds $\{42, 123\}$, trained for $50$ epochs on ViC and $20$ epochs on Y/N Vocab.
Tables~\ref{tab:repr6-vic} and~\ref{tab:repr6stv} report mean results across seeds.

\begin{table}[h]
\centering
\footnotesize
\caption{Fixed 6-dimensional representation search: ViC (mean over seeds 42, 123).}
\label{tab:repr6-vic}
\begin{tabular}{lccccc}
\toprule
Architecture & val. loss & 95\% CI & $a_{\text{mean}}$ & $a_{\text{std}}$ & $a_{\max}$ \\
\midrule
1L\_identity\_6 & 0.4308 & [0.4280, 0.4336] & 1.173 & 0.098 & 1.459 \\
3L\_relu\_16\_8\_6 & 0.4309 & [0.4279, 0.4337] & 1.254 & 0.146 & 1.614 \\
2L\_relu\_8\_6 & 0.4311 & [0.4282, 0.4338] & 1.245 & 0.152 & 1.651 \\
2L\_relu\_12\_6 & 0.4313 & [0.4284, 0.4340] & 1.245 & 0.152 & 1.649 \\
1L\_tanh\_6 & 0.4315 & [0.4285, 0.4342] & 1.193 & 0.104 & 1.435 \\
2L\_relu\_8\_6\_dropout & 0.4316 & [0.4288, 0.4343] & 1.186 & 0.116 & 1.407 \\
1L\_relu\_6$^\dagger$ & 0.4328 & [0.4300, 0.4358] & 1.230 & 0.156 & 1.570 \\
1L\_sigmoid\_6 & 0.4348 & [0.4315, 0.4376] & 1.229 & 0.114 & 1.467 \\
2L\_sigmoid\_8\_6$^\ddagger$ & 0.4616 & [0.4588, 0.4642] & 1.086 & 0.076 & 1.507 \\
\bottomrule
\end{tabular}
\end{table}

\begin{table}[h]
\centering
\footnotesize
\caption{Fixed 6-dimensional representation search: Y/N Vocab (mean over seeds 42, 123).}
\label{tab:repr6stv}
\begin{tabular}{lccccc}
\toprule
Architecture & val. loss & 95\% CI & $a_{\text{mean}}$ & $a_{\text{std}}$ & $a_{\max}$ \\
\midrule
2L\_relu\_12\_6 & 0.5058 & [0.5037, 0.5078] & 0.256 & 0.083 & 0.572 \\
1L\_relu\_6 & 0.5058 & [0.5038, 0.5078] & 0.254 & 0.085 & 0.569 \\
3L\_relu\_16\_8\_6 & 0.5060 & [0.5040, 0.5080] & 0.255 & 0.079 & 0.525 \\
2L\_relu\_8\_6 & 0.5061 & [0.5040, 0.5081] & 0.255 & 0.082 & 0.575 \\
1L\_tanh\_6 & 0.5062 & [0.5042, 0.5083] & 0.255 & 0.083 & 0.990 \\
2L\_relu\_8\_6\_dropout & 0.5083 & [0.5063, 0.5102] & 0.247 & 0.063 & 0.407 \\
1L\_identity\_6 & 0.5096 & [0.5077, 0.5117] & 0.251 & 0.086 & 1.105 \\
1L\_sigmoid\_6 & 0.5131 & [0.5110, 0.5150] & 0.241 & 0.048 & 0.434 \\
2L\_sigmoid\_8\_6$^\ddagger$ & 0.5206 & [0.5186, 0.5225] & 0.233 & 0.0005 & 0.237 \\
\bottomrule
\end{tabular}
\end{table}

\noindent $^\dagger$\texttt{1L\_relu\_6} on ViC is seed-unstable: seed $42$ gives $a_{\text{std}}=0.169$, val.\ loss $0.4343$, while seed $123$ gives $a_{\text{std}}=0.143$, val.\ loss $0.4314$.\\
$^\ddagger$\texttt{2L\_sigmoid\_8\_6} (a stacked Sigmoid$\to$Sigmoid representation) collapses to an item-independent constant $(a,b)$ on at least one seed of both datasets: on ViC, seed $42$ collapses ($a \equiv 0.949$, $a_{\text{std}} \approx 3\times10^{-7}$, val.\ loss $0.4820$) while seed $123$ trains normally ($a_{\text{std}}=0.153$, val.\ loss $0.4411$); on Y/N Vocab, \emph{both} seeds collapse ($a \equiv 0.233$, $a_{\text{std}} \approx 0.0005$), giving the worst val.\ loss in the Y/N Vocab grid.

Based on the above results, we find the following general trends.
First, fixing the representation at $6$ dimensions (rather than allowing up to $8$) costs essentially nothing: all non-degenerate architectures cluster within $\approx 0.005$ val.\ loss of one another for both datasets.
Second, the gap between ReLU and identity activations is dataset-dependent: for ViC, \texttt{1L\_identity\_6} ($0.4308$) is statistically indistinguishable from the ReLU configurations ($0.4309$--$0.4313$), but for Y/N Vocab the gap is real --- \texttt{1L\_relu\_6} / \texttt{2L\_relu\_*\_6} ($\approx 0.5060$) beat \texttt{1L\_identity\_6} ($0.5096$) by $\approx 0.0036$, outside the bootstrap confidence intervals, indicating that a ReLU layer feeding the $6$-dimensional representation is a genuine improvement for Y/N Vocab.
Third, a two-layer ReLU MLP ending in $6$ dimensions (\texttt{2L\_relu\_8\_6} / \texttt{2L\_relu\_12\_6}) is essentially tied with the single-layer ReLU network and is the best or near-best configuration for both datasets, making it a good unified default.
Fourth, stacking two squashing nonlinearities (Sigmoid $\to$ Sigmoid) immediately before the $(a,b)$ output layer performs poorly in both datasets.
Particularly, it tends to collapse the representation to a single point so that every item receives an identical $(a,b)$; this should be avoided.
Fifth, for ViC, ReLU widens the discrimination distribution ($a_{\text{std}} \approx 0.15$ versus $\approx 0.10$ for identity and tanh, $a_{\max}$ up to $1.65$ versus $\approx 1.46$), and Section~\ref{sec:vic-mlp} confirms this spread reflects a stable, frequency-correlated signal rather than noise; for Y/N Vocab, $a_{\text{std}}$ is comparable across ReLU, identity, and tanh, with only sigmoid-family architectures compressing the spread.

Overall, we recommend a two-layer ReLU MLP ending in a $6$-dimensional representation --- \texttt{Linear}$(d_{\text{in}},8) \to \text{ReLU} \to \texttt{Linear}(8,6) \to \text{ReLU} \to \texttt{Linear}(6,2)$, where $d_{\text{in}}$ is the input feature dimension ($46$ for Y/N Vocab, $15$ for ViC) --- as a unified architecture for both Y/N Vocab and ViC, with $c_{\text{real}}=0.1, c_{\text{fake}}=0.0$ for Y/N Vocab and $c=0$ for ViC.
This refines our earlier observation that linear dimension reduction is competitive with deeper networks: at a fixed $6$-dimensional representation, a shallow ReLU MLP is the best or statistically tied-for-best choice for both item types considered here.
Finally, we should remember that these embeddings will be used for adaptive testing, particularly information-based administration in the S2A3 system \cite{sharpnack2025s2a3}.
The ability for the calibration engine to distinguish between low and high discrimination is critical for effective adaptive administration, leading us to prefer architectures with higher discrimination spread such as the 2-layer ReLU.

\section{Conclusion}
\label{sec:conclusion}

We presented a neural parameterization of the 3PL IRT model in which a feature network maps item content to a low-dimensional representation $h_j = z(x_j) \in \mathbb{R}^d$, from which the discrimination and difficulty parameters $(a,b)$ are read off by a linear head, and we fit this network jointly with the latent test-taker abilities via Monte Carlo EM (Section~\ref{sec:methods}).
Evaluating under an item-split protocol that holds out entire items --- simulating the feature-only setting in which an item's parameters must be derived from content alone (Section~\ref{sec:methods-eval}) --- we asked how small and how simple this representation can be made without sacrificing held-out predictive accuracy.

The unified representation study (Section~\ref{sec:results}) gives a consistent answer across two quite different practice-test task types.
For ViC, scalar linguistic features outperform BERT-embedding-based representations at the current calibration-sample size (Section~\ref{sec:vic-embeddings}) --- the high-dimensional embeddings overfit the limited item set rather than helping --- and a single hidden layer achieves competitive test loss and discrimination range (Section~\ref{sec:vic-mlp}).
It is possible that the BERT features will improve log-loss if trained on a significantly larger item bank such as the certified DET.
For Y/N Vocab, a grid search over the guessing parameters identifies $c_{\text{real}}=0.1$, $c_{\text{fake}}=0.0$ as the best-fitting setting on item-split data (Section~\ref{sec:stv-c-grid}).
Fixing the representation dimension at $d=6$ and searching over architectures (Section~\ref{sec:repr6}), a two-layer ReLU MLP --- $\texttt{Linear}(d_{\text{in}},8) \to \text{ReLU} \to \texttt{Linear}(8,6) \to \text{ReLU} \to \texttt{Linear}(6,2)$ --- is the best or statistically tied-for-best choice for both Y/N Vocab and ViC, while several deeper or sigmoid-stacked variants collapse to a near-constant discrimination parameter and should be avoided.
This single $6$-dimensional architecture, with task-specific $c$, is our recommended unified representation $z$ for both task types.

These results demonstrate the effectiveness of MCEM to tune hyperparameters and neural net architecture for item embeddings.
The MCEM procedure used here produces point estimates of $(a,b)$ and of the feature network's weights, whereas the downstream SPICE calibration engine in the S2A3 pipeline (Figure~\ref{fig:s2a3-pipeline}) requires a full posterior over item parameters.
The natural next step is to fix $z$ at the recommended $6$-dimensional architecture, compute $h_j = z(x_j)$ for every item --- including items with no responses --- and supply these embeddings as features to SPICE, so that posterior uncertainty over $(a,b)$ can be obtained jointly with the operational item bank.
Other directions for future work include extending the representation search to additional item types beyond Y/N Vocab and ViC, evaluating whether a single $z$ can be shared across item types rather than fit separately, and revisiting the character-level and learned-$c$ extensions(Section~\ref{sec:methods-model}) now that a strong scalar baseline and representation size have been established.

\section*{Acknowledgments}

We thank Steven Nydick, Alina A. von Davier, Mancy Liao, Phoebe Mulcaire, and J. R. Lockwood for helpful discussions and feedback.

\bibliographystyle{plainnat}
\bibliography{references}

@book{lord1980,
  author    = {Frederic M. Lord},
  title     = {Applications of Item Response Theory to Practical Testing Problems},
  year      = {1980},
  publisher = {Routledge},
  address   = {Hillsdale, NJ}
}

@book{wright1979best,
  title={Best Test Design: Rasch Measurement},
  author={Wright, Benjamin D. and Stone, Mark H.},
  year={1979},
  publisher={MESA Press},
  address={Chicago, IL}
}

@article{laflair2022digital,
  title={Digital-first assessments: A security framework},
  author={LaFlair, Geoffrey T and Langenfeld, Thomas and Baig, Basim and Horie, Andr{\'e} Kenji and Attali, Yigal and von Davier, Alina A},
  journal={Journal of Computer Assisted Learning},
  volume={38},
  number={4},
  pages={1077--1086},
  year={2022},
  publisher={Wiley Online Library}
}

@article{way1998protecting,
  title={Protecting the integrity of computerized testing item pools},
  author={Way, Walter D},
  journal={Educational Measurement: Issues and Practice},
  volume={17},
  number={4},
  pages={17--27},
  year={1998},
  publisher={Wiley Online Library}
}

@article{cheng2014motivation,
  title={Motivation and test anxiety in test performance across three testing contexts: The CAEL, CET, and GEPT},
  author={Cheng, Liying and Klinger, Don and Fox, Janna and Doe, Christine and Jin, Yan and Wu, Jessica},
  journal={Tesol Quarterly},
  volume={48},
  number={2},
  pages={300--330},
  year={2014},
  publisher={Wiley Online Library}
}

@book{fischer1973lltm,
  author = {Fischer, Gerhard H.},
  title = {Log Linear Trait Models: An Approach to Item Analysis and Test Construction},
  year = {1973},
  publisher = {Psychometric Society},
  address = {Chicago, IL}
}

@inproceedings{mccarthy2021jump,
  title={Jump-starting item parameters for adaptive language tests},
  author={McCarthy, Arya D and Yancey, Kevin P and LaFlair, Geoffrey T and Egbert, Jesse and Liao, Manqian and Settles, Burr},
  booktitle={Proceedings of the 2021 conference on empirical methods in natural language processing},
  pages={883--899},
  year={2021}
}

@inproceedings{yancey2024bert,
  title={BERT-IRT: Accelerating Item Piloting with BERT Embeddings and Explainable IRT Models},
  author={Yancey, Kevin P and Runge, Andrew and Laflair, Geoffrey and Mulcaire, Phoebe},
  booktitle={Proceedings of the 19th Workshop on Innovative Use of NLP for Building Educational Applications (BEA 2024)},
  pages={428--438},
  year={2024}
}

@article{sharpnack2024autoirt,
  title={AutoIRT: Calibrating Item Response Theory Models with Automated Machine Learning},
  author={James Sharpnack and Phoebe Mulcaire and Klinton Bicknell and Geoff LaFlair and Kevin Yancey},
  journal={arXiv preprint arXiv:2409.08823},
  year={2024},
  institution={Duolingo, Pittsburgh, PA}
}

@article{sharpnack2025s2a3,
  title={{S2A3}: Thompson Sampling and Stochastic Exposure Control for High-Stakes {CATs}},
  author={James Sharpnack and Alexander Tsigler and J. R. Lockwood and Steven Nydick and Alina A. von Davier},
  journal={arXiv preprint arXiv:2606.07364},
  year={2026},
  institution={Duolingo, Pittsburgh, PA}
}

@article{nydick2026scalable,
  title={A Scalable Parametric Item Calibration Engine ({SPICE}) for Explanatory {IRT} with Sparse Data},
  author={Nydick, Steven W. and Liao, Manqian and Lockwood, J.R.},
  journal={arXiv preprint arXiv:2605.21782},
  year={2026}
}

@book{vondavier2021computational,
  title={Computational Psychometrics: New Methodologies for a New Generation of Digital Learning and Assessment},
  editor={von Davier, Alina A. and Mislevy, Robert J. and Hao, Jiangang},
  year={2021},
  publisher={Springer},
  address={Cham, Switzerland},
  series={Methodology of Educational Measurement and Assessment}
}

@article{cardwell2022duolingo,
  title={Duolingo {English} {Test}: {Technical} Manual},
  author={Cardwell, Ramsey and LaFlair, Geoffrey T. and Settles, Burr},
  journal={Duolingo Research Report},
  year={2022}
}

@article{attali2022interactive,
  title={The interactive reading task: Transformer-based automatic item generation},
  author={Attali, Yigal and Runge, Andrew and LaFlair, Geoffrey T. and Yancey, Kevin and Goodwin, Sarah and Park, Yena and von Davier, Alina A.},
  journal={Frontiers in Artificial Intelligence},
  volume={5},
  pages={903077},
  year={2022},
  publisher={Frontiers},
  doi={10.3389/frai.2022.903077}
}

@misc{davies2008word,
  author = {Davies, Mark},
  title = {{Word frequency data from The Corpus of Contemporary American English (COCA)}},
  year = {2008},
  howpublished = {\url{https://www.wordfrequency.info}}
}

@book{councilofeurope2001,
  title     = {Common European Framework of Reference for Languages: Learning, Teaching, Assessment},
  author    = {{Council of Europe}},
  year      = {2001},
  publisher = {Cambridge University Press},
  address   = {Cambridge}
}

@inproceedings{kingma2014adam,
  title={Adam: A Method for Stochastic Optimization},
  author={Kingma, Diederik P. and Ba, Jimmy},
  booktitle={International Conference on Learning Representations (ICLR)},
  year={2015}
}

\end{document}